\documentclass{aa}  

\usepackage{graphicx}
\usepackage{txfonts}
\usepackage{bm}
\usepackage{subcaption}
\usepackage{xcolor}

\usepackage{placeins}

\newcommand{\rmm}[1]{\xspace} 

\usepackage{natbib,twoopt}
\usepackage[breaklinks=true]{hyperref} 

\begin{document} 

   \title{How do supernova remnants cool?}
   \subtitle{II. Machine learning analysis of supernova remnant simulations}

   \author{P. Smirnova\inst{1}\thanks{smirnova@ph1.uni-koeln.de},
          E. I. Makarenko\inst{1}\thanks{makarenko@ph1.uni-koeln.de},
          S. D. Clarke\inst{2},
          E. Glukhov\inst{3},
          S. Walch\inst{1,4},
          I. Vaezzadeh\inst{1},
          D. Seifried\inst{1}
          }

   \institute{I. Physikalisches Institut, Universit{\"a}t zu K{\"o}ln, Z{\"u}lpicher Str. 77, D-50937 K{\"o}ln, Germany
    \and Institute of Astronomy and Astrophysics, Academia Sinica, No. 1, Sec. 4, Roosevelt Rd., Taipei 10617, Taiwan, 
    \and Stony Brook University, 100 Nicolls Rd, Stony Brook, NY 11794, USA, 
    \and Center for Data \& Simulation science, Universit{\"a}t zu K{\"o}ln, Albertus-Magnus-Platz, D-50923 K{\"o}ln, Germany}

   \date{Received 17 June 2024 / Accepted 16 November 2024}

  \abstract
    {About 15\%-60\% of all supernova remnants are estimated to interact with dense molecular clouds.
    In these high density environments, radiative losses are significant.
    The cooling radiation can be observed in forbidden lines at optical wavelengths.}
    {We aim to determine whether supernovae at different positions within a molecular cloud (with/without magnetic fields) can be distinguished based on their optical emission, e.g. H$\alpha$~($\lambda\,6563$), H$\beta$~($\lambda\,4861)$, [\ion{O}{iii}]~($\lambda\,5007$), [\ion{S}{ii}]~($\lambda\,6717, 6731)$, [\ion{N}{ii}]~($\lambda\,6583)$, using machine learning (e.g.~principle component analysis and k-means clustering).}
    {We have conducted a statistical analysis of the optical line emission of simulated supernovae interacting with molecular clouds that formed from the multi-phase interstellar medium modelled in the SILCC-Zoom simulations with and without magnetic fields.
    This work is based on the post-processing of simulations which have been carried out with the 3-D (magneto)hydrodynamic code \textsc{FLASH}.
    Our data set consists of 22 simulations.
    The supernovae are placed at a distance of either 25\,pc or 50\,pc from the molecular cloud centre of mass.
    First, we calculate optical synthetic emission maps (taking into account dust attenuation within the simulation sub-cube) with a post-processing code based on MAPPINGS V cooling tables.
    Second, we analyse the data set of synthetic observations using principle component analysis to identify clusters with the k-means algorithm.
    In addition, we make use of BPT diagrams as a diagnostic of shock-dominated regions.}
   {We find that the presence or absence of magnetic fields has no statistically significant effect on the optical line emission.
    However, the ambient density distribution at the site of the supernova changes the entire evolution and morphology of the supernova remnant.
    Due to the different ambient densities in the 25\,pc and 50\,pc simulations, we are able to distinguish them in a statistically significant manner.
    Although, optical line attenuation within the supernova remnant can mimic this result depending on the attenuation model that is used.
    That is why, multi-dimensional analysis of optical emission line ratios in this work does not give extra information about the environmental conditions (ambient density and ambient magnetic field) of the SNR.} 
  {}
  

   \maketitle
%

\section{Introduction}

Typically, massive stars of 8 to 40\,M$_{\sun}$ explode as a type II supernova (SN) at the end of their lifetime.
Around 15\%-60\% of these SNe are estimated to interact with a nearby molecular cloud (MC) \citep{Hewitt2009, Zhou2023}.
Interaction with the dense environment of a MC shapes the evolution of the SN remnant (SNR).
First, the SN explodes, producing a medium mostly filled with hot gas \citep{McKee1977, Kavanagh2013, Alsabti2017}.
The ejecta undergoes free expansion, in which the shocks produced by the explosion significantly affect the gas, which becomes hot, ionised, and turbulent.
In the next step, it evolves to the Sedov-Taylor (adiabatic blast wave) stage and the ejecta energy is transferred to the ambient gas, but the net energy is conserved \citep{Sedov1959, Truelove1999, Haid2016}.
At this stage (typically around $10^{4}$ years after the explosion, but the exact number depends on the ambient density distribution), the young SNR can be observed in the X-ray and even $\gamma$-ray regimes \citep{Borkowski2001,Aharonian2004, Vink2012, Sasaki2012, Slane2014}.
The third stage starts when radiative losses become significant, at which point a shell-like structure appears.
Due to the cooling, the remnant starts to emit at UV and optical wavelengths, while further expanding into (snow-ploughing through) the ambient medium \citep{Fesen1985, Mavromatakis2002, Boumis2008, Fesen2024}.
About 20\% of the SNRs in our Galaxy have such optical counterparts \citep{Green19}.
In the final stage, the remnant dissolves in the interstellar medium (ISM) \citep{Ostriker1988}.

The evolution of a SN in a homogeneous medium has already been well studied \citep{McKee1977, Cioffi1988, Haid2016, Jimenez2019}, but in reality, SNe are immersed in a complex ISM - a highly inhomogeneous environment with a wide range of densities and temperatures.
Therefore, simulations of MCs are essential to our comprehension of the structure of the ISM, and as a result to modelling realistic SNRs.
Recently, great progress has been made in simulations of the complex multi-phase ISM and the interaction of SNe with the turbulent ISM \citep[e.g.][]{deAvillez&Breitschwerdt2005, Gatto2015, Walch2015, Walch&Naab2015, Zhang2019, Haid2019, Ganguly2023} or with MCs in particular \citep{Iffrig15, Seifried2017, Seifried2018}. 
All of these simulations help to answer many physical questions, e.g.~the effect of SNe on the star formation rate \citep{Padoan2011A, Gatto2015}, and the energy and momentum contribution of SNe to the ISM \citep{Walch&Naab2015}.

\citet{Makarenko2023} studied optical emission (H$\alpha$~($\lambda\,6563$), H$\beta$~($\lambda\,4861)$, [\ion{O}{iii}]~($\lambda\,5007$), [\ion{S}{ii}]~($\lambda\,6717, 6731)$, [\ion{N}{ii}]~($\lambda\,6583)$) from SNRs, employing the post-processing module CESS (Cooling Emission in the optical band from Supernovae in (M)HD Simulations)\footnote{\href{CESS: Cooling Emission in the optical band from Supernovae in (M)HD Simulations}{https://github.com/kativmak/CESS}} used for the \textsc{FLASH} code.
We employ the updated collision data output from \textsc{MAPPINGS V} \citep{Sutherland2017}.
\citet{Makarenko2023} show that it is crucial to consider both the attenuation effect due to the dust within the SNR bubble using a simple radiative transfer, and a realistic density distribution from the surrounding ISM, as this significantly affects the line emission and hence the diagnostics based on different lines.
Correspondingly, the more complex simulations we have, the more difficult it is to disentangle and study the impact of each parameter (e.g.,~temperature, density, magnetic field) on the simulation outcome using non-statistical methods.

With the rapid development of supercomputers (and simulations), and an increasing number of telescopes (and accumulated observations), large data sets are now common in astrophysics \citep[e.g.][]{Hipparcos1997, SDSS2022, Gaia2021}.
Because of this, the search for relations between physical parameters can be hampered by the large amount of data, and indeed correlations can be found between more than two parameters.
In this case, tools such as unsupervised machine learning can be extremely useful: one such tool which is suitable for searching for correlations between a large number of parameters is principal component analysis (PCA).
PCA transforms a large set of variables to a smaller one which still contains most of the information from the larger set, losing a little accuracy for simplicity.
\citet{Einasto2011} presented a method which uses PCA to investigate the strength of correlations between the properties of superclusters of galaxies (data from SDSS DR7) and search for the presence of distance-dependent selection effects in the supercluster catalogue.
PCA can also be applied for dimension reduction or to visualise data \citep{DRPCA}.
Several other algorithms could also be applied, such as t-distributed stochastic neighbour embedding (t-SNE), as used by \citet{Anders2018}, to better distinguish chemical sub-populations in the solar vicinity, rather than looking at 2-D abundance maps.
Once the dimensions are reduced, it is important to robustly find clusters with unsupervised methods.

The k-means algorithm \citep{MacQueen1967, Hartigan1979} searches for proximity in multi-dimensional space (in our case in the reduced dimensional space).
For example, the algorithm was used in \citet{Rubin2016} to divide light curves into classes with a fixed number of clusters.
The Sloan Digital Sky Survey (SDSS) used the k-means algorithm to identify clusters of different stellar spectral classes as well as rare objects and outliers \citep{Sanchez2013}.

Further, there have been many attempts recently to improve the classification of supernovae in BPT (Baldwin, Phillips \& Terlevich) diagrams \citep[][]{Baldwin81, Kauffmann2003, Kewley2019_diagn} since the classification of different astrophysical objects is not always unambiguous or does not reflect all physical features.
BPT diagrams allow the determination of the main excitation source of an object (shocks or photo-ionisation) using optical line ratios.
The idea to use multi-dimensional data classification for emission-line galaxies with the support vector machine algorithms has been tried both for observations \citep{Stampoulis2019} and for theoretical models \citep{Kopsacheili2020}.
There have been several attempts to improve the BPT diagrams using other optical line ratios (creating a multi-dimensional optical line ratio space), and modern machine learning techniques to better classify observed objects or to better constrain their physical conditions \citep[e.g.][]{Vogt2014, Ho2019, Zhang2020, Ji2020, Rhea2021}.
However, none of the new models constitute a universal tool that could be used in both theoretical work and observations.
Here we test whether BPT diagrams are a sensitive diagnostic tool to uncover the environmental conditions of young SNRs.

In this paper, we perform a statistical study using unsupervised machine learning by post-processing (M)HD simulations, comparing different initial conditions such as magnetic field strength and distances of the SN explosion to the centre of mass of the nearby MC which the remnant interacts with.
We use multi-dimensional data consisting of multiple optical line ratios. 
The emission lines are calculated by post-processing the 3-D simulations with the \textsc{MAPPINGS V} code.
Our analysis reveals whether line ratios are sensitive to variations in the pre-shock density distribution and the presence of the magnetic field at the site of the SNR.

The paper is structured as follows.
In Section~\ref{sec:data set} we describe the simulation setups, as well as the post-processing routine used to calculate optical emission lines.
Section~\ref{sec:stat_analyse} describes the statistical methods (normalisation, pre-processing and clustering) applied to our data and BPT diagrams.
We present our results in Section~\ref{sec:results} and discuss the importance of the ambient ISM properties and the attenuation effect for the optical emission of SNR.
The conclusions are given in Section~\ref{sec:conclusions}.

\section{Data set} \label{sec:data set}

\begin{figure}
    \centering
    \includegraphics[width=\columnwidth]{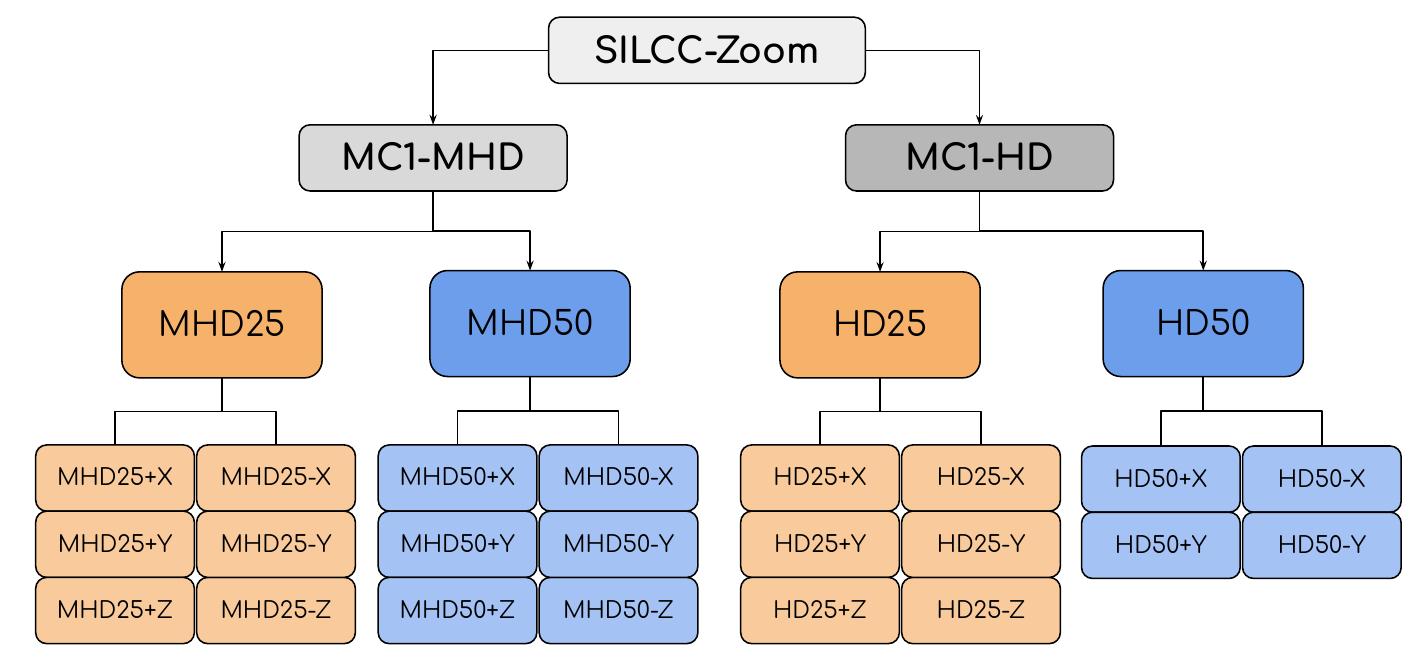}
    \caption{Hierarchical structure of the data set from the SILCC-Zoom project. We have in total 22 simulations: "MC1-MHD" (with a magnetic field) and "MC1-HD" (without a magnetic field). Each of the "MD25", "MHD50", "HD25, "HD50" data sets contains all possible positions of SN event ($\pm$X, $\pm$Y, $\pm$Z). Further details can be found in \cite{Seifried2018}.}
    \label{fig:data set}
\end{figure}

We use SILCC-Zoom simulations of SNe interacting with MCs from \cite{Seifried2018} using the \textsc{FLASH} code \citep{Fryxell2000} as an initial data set.
The data set description can be found in Fig.~\ref{fig:data set}; for more details, see \cite{Walch2015, Girichidis2016, Seifried2017, Seifried2018}.
In Section~\ref{sec:simulations}, we briefly summarise the simulation setup.
In Section~\ref{sec:opt_emission} we describe the post-processing tool from \cite{Makarenko2020, Makarenko2023} which is used to produce optical emission cubes and maps.

\subsection{Simulations}\label{sec:simulations}
\begin{table}
    \caption{Main parameters of MCs. Name of the simulation data set (first column), total mass (first column), considered sub-volume volume (second column), and molecular hydrogen fraction within the selected sub-volume (third column).}
    \centering
    \begin{tabular}{c c c c }
         \hline \hline
         Simulation & Total mass & Volume & H$_2$ fraction\\
         & [M${_\odot}]$ & [pc$^3$] & [M$_\odot$] \\
         \hline
         MC1-HD & 7.3 $\times$ 10$^4$ & 88 $\times$ 78 $\times$ 71 & 2.1 $\times$ 10$^4$\\
         MC1-MHD & 7.8 $\times$ 10$^4$ & 88 $\times$ 78 $\times$ 71 & 1.3 $\times$ 10$^4$\\
         \hline
    \end{tabular}
    \label{tab:mcs_params}
\end{table}

\begin{figure*}
\includegraphics[clip,width=19cm]{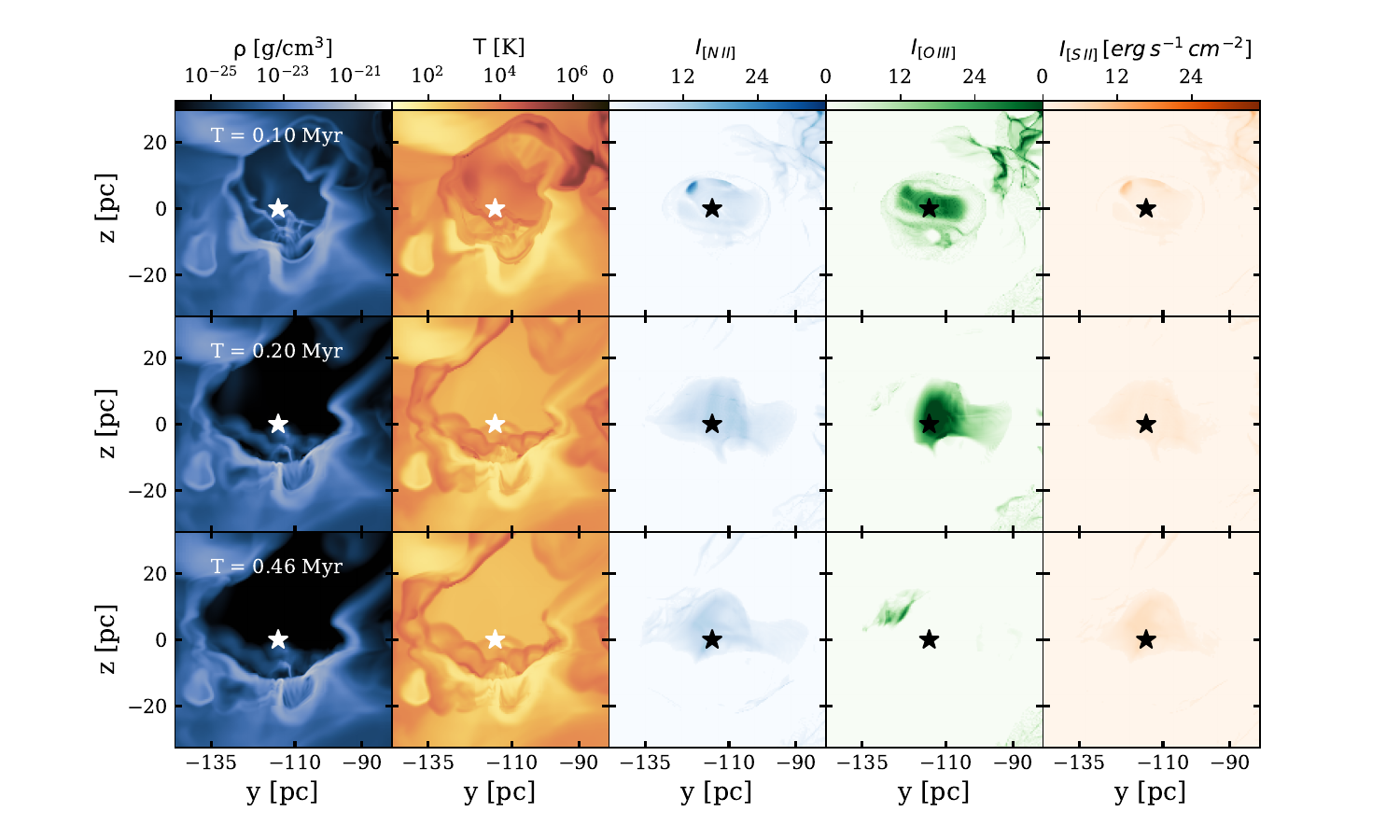}
    \caption{Time evolution (from top to bottom) of (from left to right): density slice, temperature slice, [\ion{N}{ii}]~($\lambda$\,6583Å) intensity projection, [\ion{O}{iii}]~($\lambda$\,5007\,Å) intensity projection, [\ion{S}{ii}]~($\lambda$\,6731\,Å) intensity projections for "MHD50+X" simulation. The star symbol in each panel represents the position of the SN explosion. The SN explosion disrupts part of the dense MC, as can be seen from the density slice (first column). This MC and SN interaction gives rise to the optical emission from the SNR. From the temperature evolution of the SNR bubble (second column), we can see the hot gas (10$^{6-7}$\,K) cooling down over 0.46\,Myr, and forming the complex structures at the edge of the SNR and MC that are subsequently visible in the optical emission. Among three optical lines (third to fifth columns), the [\ion{O}{iii}] is the strongest as it is a "volume filling emission". [\ion{N}{ii}] and [\ion{S}{ii}] originate from a thin cooling layer right behind the shock front. 
    }
    \label{fig:temp_dens}
\end{figure*}

In the SILCC project \citep{Walch2015, Girichidis2016}, we model the evolution of a section of a galactic disc (500\,pc $\times$ 500\,pc $\times$ $\pm$ 5\,kpc), with an initial gas surface density of $\Sigma_{\mathrm{gas}}=10\,\mathrm{M_{\odot}\,pc}^{-2}$ with initial conditions similar to the ones present in the solar neighbourhood.
The simulations take into account SN feedback, magnetic fields (in MHD runs, $B_{\mathrm 0,x}\sim$~4\,$\mu$G is the initial field strength in the disc midplane) and self-gravity with a non-equilibrium treatment of the H$_2$ and CO chemistry of the gas.
The rate of SNe is in agreement with the Kennicutt–Schmidt relation for our surface density \citep{Kennicutt1998} (15\,Myr$^{-1}$).
The SNe explosions heat and stir the gas and a complex multi-phase ISM appears.
A zoom-in strategy is used to resolve the formation of MCs to scales of $\sim 0.12$\,pc in the SILCC-Zoom simulations \citep{Seifried2017, Seifried2018}.
We select two basic simulations: one hydrodynamical (HD) without magnetic fields and one magnetohydrodynamical (MHD) with magnetic fields: MC1-HD and MC1-MHD.
The general properties of the MCs which we will study further are given in Table~\ref{tab:mcs_params}.

After the formation of MCs (we denote this time as $t_{\rm 0}$), and some further evolution, a new SN is exploded at time $t_{\rm SN} = t_{\rm 0}$ + 1.53\,Myr.
In \cite{Seifried2018} a new simulation is run for each SN exploding at different distances (25\,pc or 50\,pc simulations are used here), and at positions with respect to the centre of mass of the cloud along the $x, y$ and $z$-directions in a zoom-in region (see Fig.~\ref{fig:data set}).
Each SN explosion is modelled by adding 10$^{51}$\,erg of thermal energy into the radius $R_{\rm inj}$ around the explosion centre.
We make sure to resolve the Sedov-Taylor radius with at least 4 simulation cells, corresponding to 0.48\,pc \citep{Gatto2015}.
We follow each SNR evolution for 0.46\,Myr. 
It is interesting to consider different SN positions because of the variation in density distribution at the site of the SN explosion (10$^{-27.5}$~--~10$^{-21.5}$\,g\,cm$^{-3}$ at 25\,pc, and 10$^{-28}$~--~10$^{-21}$\,g\,cm$^{-3}$ at 50\,pc).
As a result, the optical emission, which arises from interactions with a dense medium, will also look different.
The aim of this paper is to explore how this difference is manifested.

Examples of the time evolution of temperature and density are shown in Fig.~\ref{fig:temp_dens}, from top to bottom in columns 1 and 2 respectively.
As seen in the upper right corner of the temperature evolution, the SN blows out the gas from the cavity towards the region of lower-density.
During this process, the gas cools down and starts to emit in the optical ([\ion{N}{ii}], [\ion{O}{iii}] and [\ion{S}{ii}] lines; third, fourth and fifth columns of Fig.~\ref{fig:temp_dens} correspondingly).
At the end of the evolution ($t = 0.46$\,Myr), the optical emission fades away, as the majority of the SNR bubble is cooled below typical optical emission temperatures (less than a few $10^4$\,K).
Because the different parts of the bubble evolve on different timescales (due to the complex shock-cloud interaction), we can still see optical emission at $t = 0.46$\,Myr in the central region and in the upper-right part of the SNR bubble.

\subsection{Optical emission post-processing}\label{sec:opt_emission}
\begin{table}
    \caption{The median percentage of attenuation for each line with the error (columns 2-6) in each data set (column 1). The attenuation of each optical line is taken into account as described in Section~\ref{sec:opt_emission}.}
    \label{tab:lines_atenuation}
    \centering
    \begin{tabular}{c c c c c c}
          \hline \hline
          data set & \multicolumn{5}{c}{Median \% of the line attenuation}\\
          & [\ion{S}{ii}] & [\ion{N}{ii}] & [\ion{O}{iii}] & H$\alpha$ & H$\beta$ \\
         \hline
         MHD25 & 18 $\pm 10$& 25 $\pm 9$& 39 $\pm 10$& 32 $\pm 10$& 19 $\pm 11$ \\
         HD25 & 34 $\pm 5$& 39 $\pm 5$& 40 $\pm 7$& 40 $\pm 5$& 34 $\pm 5$ \\
         MHD50 & 34 $\pm 14$& 38 $\pm 12$& 43 $\pm 12$& 39 $\pm 10$ & 35 $\pm 14$ \\
         HD50 & 37 $\pm 9$& 36 $\pm 10$& 34 $\pm 14$ & 33 $\pm 10$ & 28 $\pm 13$ \\
         \hline
    \end{tabular}
\end{table}

To prepare our simulations and reduce the computational cost of post-processing,
we cut out sub-cubes using the biggest SNR bubble radius as a border (which can be measured at $t = 0.46$\,Myr).
Each sub-cube is 64.8\,pc on a side.
In this case, we take into account the emission mainly from the SNR and avoid significant contamination from the background (previous SN events) in our analysis.

We follow the method of \citet{Makarenko2023}, which introduces the post-processing module CESS for the \textsc{FLASH} code, which uses the collision data from \textsc{MAPPINGS V} \citep{Sutherland2017, mappings_code} to reproduce optical emission maps of simulated SNRs.

In summary, the post-processing procedure is as follows. First, from the .hdf5 simulation output we cut out a sub-cube of the region containing the SNR (64.8\,pc on a side), and re-grid the AMR grid to a uniform grid (every cell is 0.12\,pc$^3$).
Then, we calculate the emitted luminosity for every cell in the 3-dimensional computational domain using temperature-cooling rate dependency, known from \textsc{MAPPINGS V}.
After that, we integrate along a given line of sight taking into account the attenuation (projecting the 3-D cube to a 2-D map):

\begin{equation}
    F_{\rm tot} = \int F_i e^{- \tau_{\rm i}} \,ds 
\end{equation}
where $F_{\rm i}$ is the flux of the cell $i$, $\tau_{\rm i}$ is the optical depth, $ds$ is the area of the cube.
The optical depth $\tau_{\rm i}$ is calculated in the following manner:
\begin{equation}
    \tau_i = \kappa_{\rm abs} \rho_i V_i^{1/3} f_{\rm d},
\end{equation}
where $\kappa_{\rm abs}$ is the dust absorption cross section per mass of dust ($\mathrm{cm^{2} g^{-1}}$), $\rm \rho_i$ is the density of cell $i$, $V_i$ is the cell volume, and $f_{\rm d}$ is the dust-to-gas ratio ($f_{\rm d} = 0.01$ is fixed in our simulations).
The dust absorption cross-section is taken from \citet{WeingartnerDraine2001} (Milky Way dust with $\mathrm{R_V} = 4.0$). The attenuation percentage for each line in our data set can be found in Table~\ref{tab:lines_atenuation}.
The attenuation effect can be high for the optical emission lines due to gas and dust attenuation. 
Here we take into account only attenuation within the SNR simulation domain.
This attenuation effect leads to fewer SNRs detected in the optical regime in our Galaxy as compared to radio and X-ray \citep{Green19}.
Thus, we reproduce optical emission line maps ($\mathrm{H\alpha}$~($\lambda\,6563\,$Å), $\mathrm{H\beta}$~($\lambda\,4861\,$Å), [\ion{O}{iii}]~($\lambda\,5007\,$Å), [\ion{S}{ii}]~($\lambda\,6717\,$Å, $6731\,$Å), [\ion{N}{ii}]~($\lambda\,6583$\,Å) that show the same features as in real observations of SNRs interacting with the dense medium.
We investigate the following optical line ratios: [\ion{O}{iii}]~($\lambda\,5007)/ \mathrm{H\alpha}$,
[\ion{S}{ii}]~($\lambda\,6731$)/$\mathrm{H\alpha}$, [\ion{N}{ii}]~($\lambda\,6583$)/$\mathrm{H\alpha}$, [\ion{O}{iii}]~($\lambda\,5007$)/$\mathrm{H\beta}$.
These line ratios are typically used in BPT diagrams as shown in Section~\ref{sec:bpt}.
As a result, we have a 4-dimensional data set.
We calculate line ratios at each timestep (0.02\,Myr) during the SNR evolution time (0.46\,Myr) along the $x$-axis from one side of the cube in 22 simulations (see Fig.~\ref{fig:data set}).

\section{Statistical analysis}\label{sec:stat_analyse}
\subsection{Data normalisation}
In order to make a comparison between data points, the data set must be normalised according to the chosen transformation function.
We would like to have the normalised variables in units relative to the standard deviation of the sample.
To get the normalised variables, we performed the following normalisation for each data point:
$$z = \frac{d - \mu}{\sigma} ,$$
where $d$ is initial (raw) data, $\mu$ is the mean of the sample, and $\sigma$ is the standard deviation.
This is standard practice for PCA to remove the effects of different means and scales between the features.
As an input for this procedure, we take optical line ratios in a linear space.

\subsection{Dimension reduction: Principal component analysis (PCA)}
\begin{table*}
    \caption{Loading for principal axes (PC1 or PC2, second column) in feature space, representing the directions of maximum variance in the data set for each line ratio (3-6 columns). The variance captured by PC1 and PC2 is spread across multiple features, indicating that those features contribute together to the variance in a similar manner.}
    \label{tab:PC1_PC2_combination}
    \centering
    \begin{tabular}{c c c c c c}
          \hline \hline
          data set & PC & \multicolumn{4}{c}{Coefficients of the line ratio}\\
          && [\ion{S}{ii}]/H$\alpha$ & [\ion{N}{ii}]/H$\alpha$ & [\ion{O}{iii}]/H$\alpha$ & [\ion{O}{iii}]/H$\beta$ \\
         \hline
         MC1$_{\rm att}$ (all data) &PC1&-0.62&0.26&0.36&0.39\\
         &PC2& 0.12&-0.52&0.70&0.70\\
         \hline
    \end{tabular}
\end{table*}

We have four optical line ratios (see Section~\ref{sec:opt_emission}) and their time evolution over 0.46\,Myr.
This yields a 4-dimensional data set.
We first reduce our data set to 2-D, to aid in data interpretation while retaining as much information as possible. 
For 3-D see Appendix~\ref{C:3DPCA}, Fig.~\ref{fig:example_pca_3D_25-50}; the third dimension contributes less than 10\% significance, therefore using 2-D does not lead to lower accuracy.
Principal components (PCs) are directions in feature space along which the original data exhibits the greatest variance.
By retaining the two PCs with the largest variances, we minimise the loss of information during dimensionality reduction.
The larger the variance of the PC axis, the larger the dispersion of the data along it (the more information it carries), therefore the less information is lost.
This then enables us to visualise the data in two dimensions.

The loading represents the weight (or contribution) of a specific original variable to the PC.
The loadings for each line ratio are shown in Table~\ref{tab:PC1_PC2_combination} for each PC of the data set.
Despite PC1 and PC2 being represented by a mixture of features, we focus on the ones with the highest variance.
It is easy to see that for PC1 the line ratio [\ion{S}{ii}]/H$\alpha$ is the most significant.
S$^{+}$ is a known tracer of SNRs, as it is a collisionally excited ion that can form in a large recombination zone behind the SNR shock.
For PC2 it is [\ion{O}{iii}]/H$\alpha$ and [\ion{O}{iii}]/H$\beta$ that are the most significant.
O$^{++}$ is forming within a significant region of the SNR bubble, as it has a higher excitation temperature than S$^{+}$ or N$^{+}$.
Due to this, the observed area f the SNR bubble, where we can observe O$^{++}$, differs in various simulations, and can be used as a parameter to distinguish them.
The result of the PCA algorithm is shown in Fig.~\ref{fig:all_data_2or4clust} and Appendix~\ref{A:PCA_for_MF}, Fig.~\ref{fig:MHDvsHD_PCA}.
Note that the data in the figures is already clustered (circles and crosses).
We describe the clustering process in Section~\ref{subsec:k-means}.

Apart from the PCA, we also try to reduce the dimensionality of the data set using the t-SNE algorithm (see Appendix~\ref{B:t-SNE}, Fig.\ref{fig:TSNE}).
This is a non-linear algorithm, in contrast to PCA.
We found no qualitative difference between its results and those from PCA, so in the following analysis, we will only use the PCA algorithm.

\subsection{Clustering: k-means and Rand index} \label{subsec:k-means}
\begin{figure}	
    \centering
    \includegraphics[width=0.8\columnwidth]{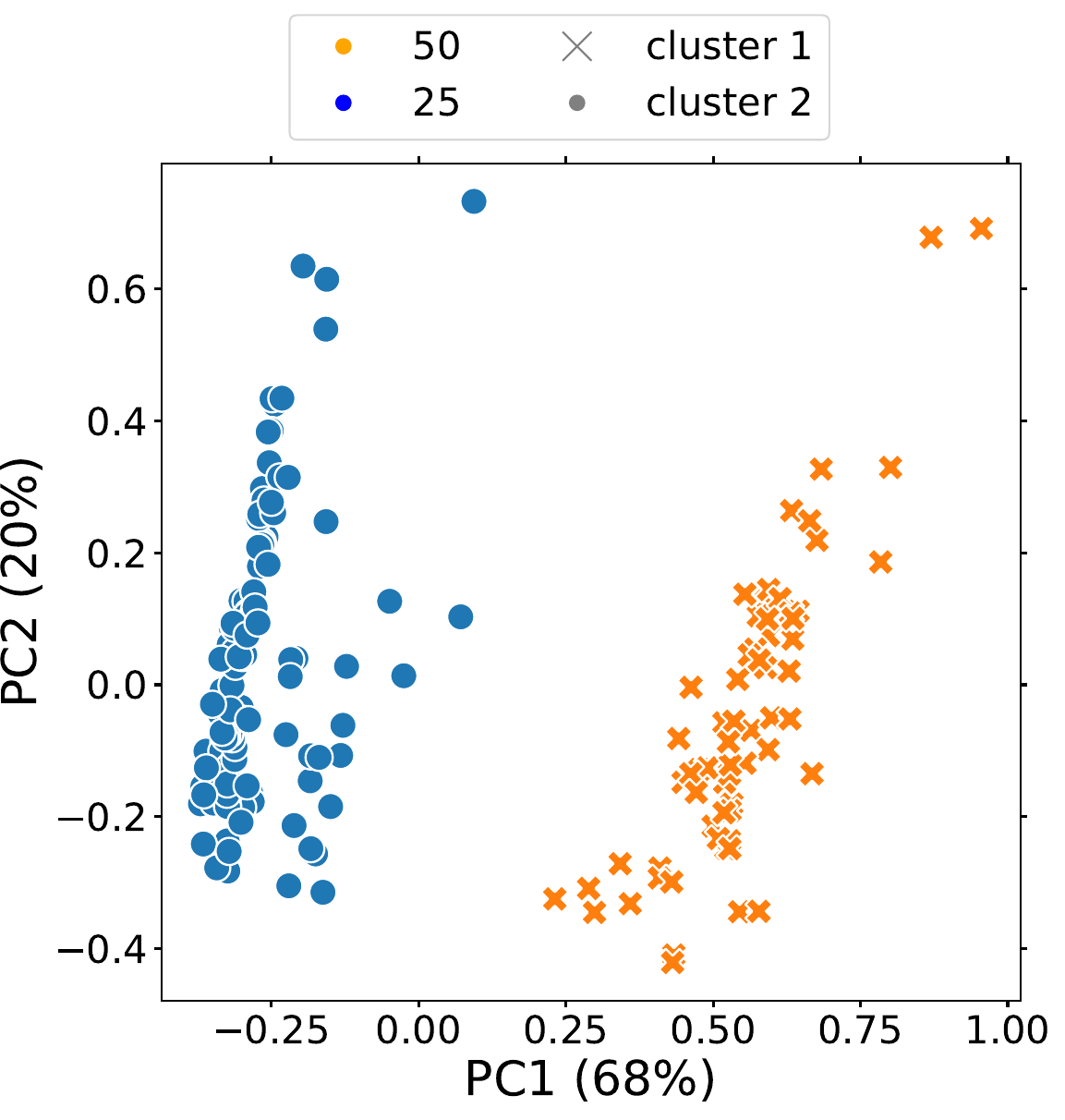}
    \caption{"MHD25", "MHD50", "HD25", "HD50" data sets (e.g. "MC1$_{\rm att}$") after using the PCA algorithm. "MHD25"+"HD25" is in orange colour, "MHD50"+"HD50" is in blue. The predicted clusters are marked with circles and crosses. The higher the percentage for each PC the higher the relative variance in the data set that is observed in the direction of the corresponding eigenvector (for the absolute values see Table~\ref{tab:PC1_PC2_combination}). We use $n_{\rm clust} = 2$ in the k-means algorithm. The data is clearly separated into two distinct clusters.}
    \label{fig:all_data_2or4clust}
\end{figure}

\begin{figure}	
\includegraphics[width=\columnwidth]{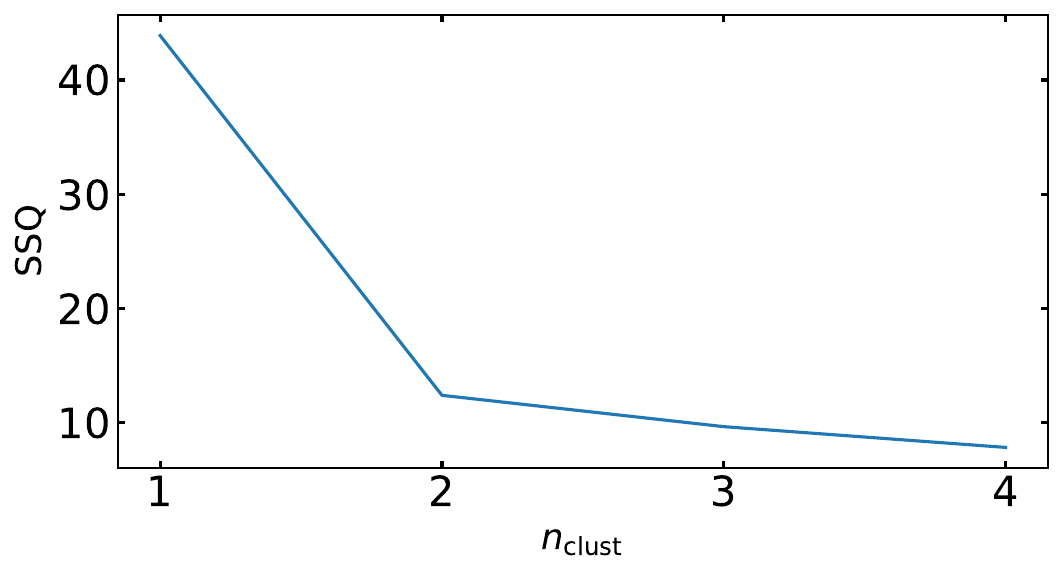}
\caption{Elbow plot for all data set ("MC1$_{\rm att}$"). On the $x$-axis the number of clusters is presented, while on the $y$-axis the total within the sum of squared distances is plotted. We see a kink in the line at $n_{\rm clust}=2$, which indicates that the optimal number of clusters to be used by the k-means algorithm is 2.}
\label{fig:elbowplot}
\end{figure}

\begin{figure}	
	\includegraphics[width=\columnwidth]{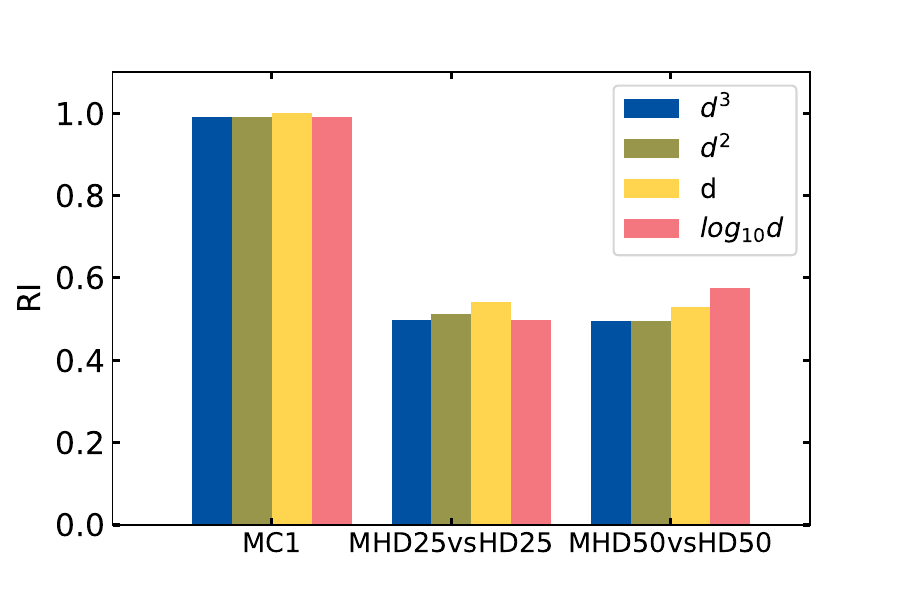}
    \caption{Rand index value on the $y$-axis for different data sets on the $x$-axis: all data try to define 2 clusters ("MC1", first column), "MHD25" vs "HD25" (second column), "MHD50" vs "HD50" (third column). The colours show the different data transformations: polynomial with blue and green, standard with yellow, and logarithm with red (see the legend for details). The higher the Rand index the more effective the clustering. The best results were obtained with standard and polynomial data for the first column (all data, distance difference).}
    \label{fig:randt_ind}
\end{figure}

After reducing the dimensionality of the data set, we group the points in 2-D space into clusters using the k-means algorithm.
This algorithm seeks to minimise the sum of the squared Euclidean distances between each point and the centroid of the cluster to which it has been assigned.
For the k-means algorithm, we need to provide the number of expected clusters as an input parameter.
The elbow method \citep{elbow_plot} is a technique to determine the number of clusters ($n_{\rm clust}$) to use in the k-means clustering algorithm.
First, the elbow method calculates the sum of the squares of the distances between all points and then calculates the mean.
When we choose $n_{\rm clust} = 1$, the sum of the squared distances within the cluster is the largest.
As the value of $n_{\rm clust}$ increases, the sum of squared distances (SSQ) within the cluster decreases.
Finally, to determine the best choice of $n_{\rm clust}$, we plot $n_{\rm clust}$ versus the SSQ as shown in Fig.~\ref{fig:elbowplot}.
At the point $n_{\rm clust} = 2$, the SSQ decreases dramatically, or forms an `elbow'.
This point is considered the optimal value of the number of clusters.
In Fig.~\ref{fig:all_data_2or4clust} the colours represent the real data set, and the symbols represent the identified clusters after performing the k-means algorithm (and vice versa in Fig.~\ref{fig:MHDvsHD_PCA}).

To verify the clusters we identify with the k-means algorithm, we can use the Rand index.
The Rand index indicates the similarity between the given (original) clusters and the predicted clusters.
It is defined as $\mathrm{RI} = \frac{a}{b},$ where $a$ is the number of agreeing pairs (original clusters vs predicted clusters), and $b$ is the total number of pairs.
The closer the Rand index is to 1.0, the better the clustering.
Fig.~\ref{fig:randt_ind} shows the Rand index for different samples.
For example, there were 198 data points in the "MC1$_{\rm att}$" data set, and for "MHD25"+"HD25", 98 data points were identified correctly to the clusters (initial and predicted point coincided), which leads us to RI = 98/198 $\sim$ 0.5.
Various transformations of the initial data (polynomial - d$^2$ or d$^3$, unscaled - d, and logarithmic - $\log_{10} d$) were also considered before proceeding with the PCA so as not to bias the result.
Despite this, the best Rand index is obtained for the unscaled data, for the data set "MC1$_{\rm att}$".
In this case, we can almost perfectly identify the "MHD25"+"HD25" and "MHD50"+"HD50" groups again after the statistical analysis if we compare the resulting clusters from the k-means algorithm with the initial data set.
Note that we do not use the information about labels for the initial data set in our analysis.
The Rand index for "MHD50+HD50" and "MHD25+HD25" is $\sim0.5$, which means that we cannot distinguish different clusters for simulations with and without a magnetic field.

\subsection{BPT diagram}\label{sec:bpt}
\begin{figure}%
    \includegraphics[trim={0.1cm 0.1cm 0.1cm 0.1cm},clip,width=\columnwidth]{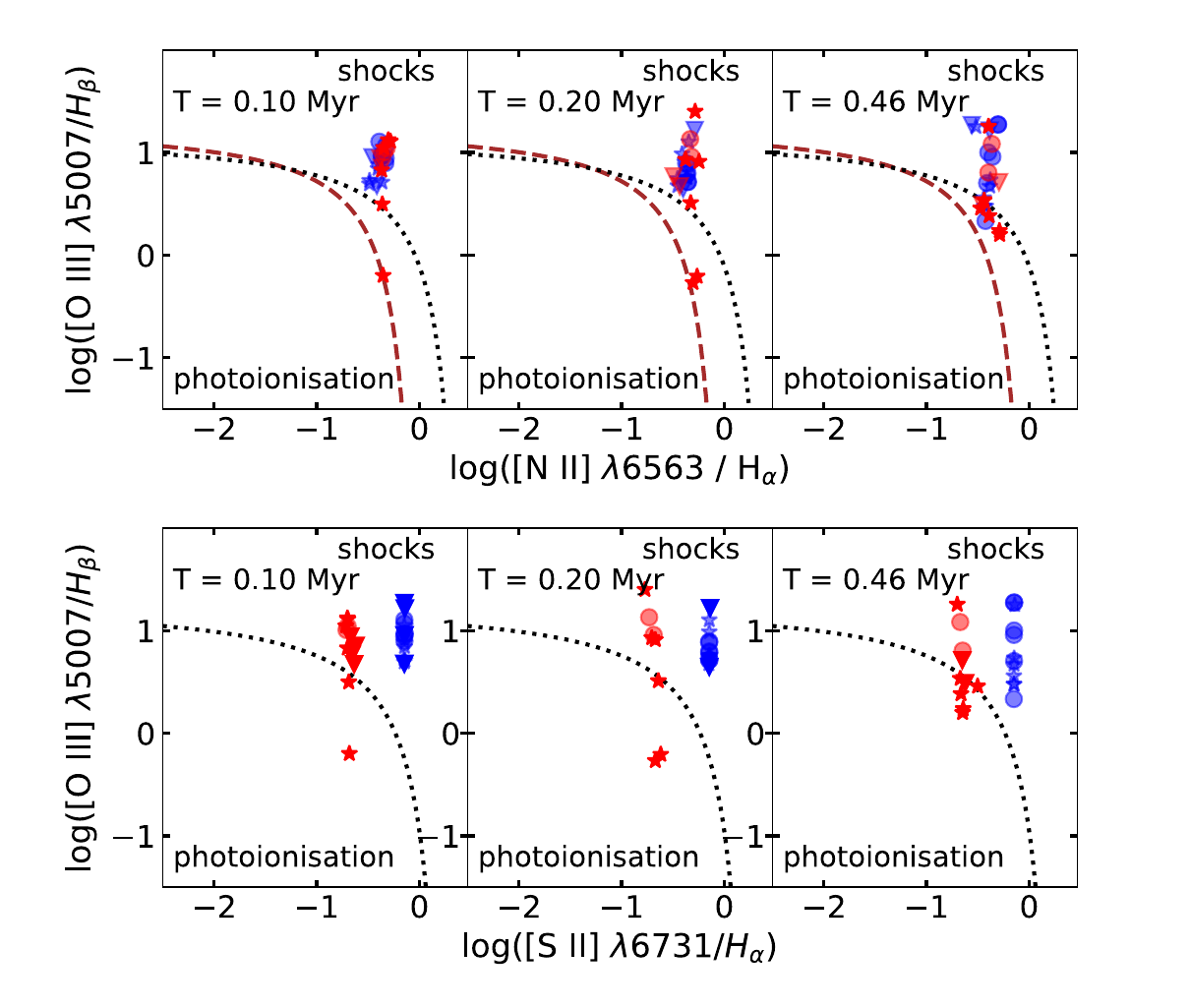}
    \caption{Time evolution (from left to right) of the classical BPT diagram (upper row) and sulphur BPT diagram (lower row) for "MHD25", "HD25" (blue) and "MHD50", "HD50" (red). Star symbols show MHD simulations and circle symbols show HD simulations. The dotted reference line is \cite{Kewley2001}, and the dashed line is from \cite{Kauffmann2003} for the upper row. The dotted reference line for the lower row is from \cite{Kewley2001}. For both figures, reference lines separate the photoionisation region (star-forming, lower left corner) and the shock-dominated region (upper right corner). SNRs in the upper figure are typically located in the "mixed region" (between the reference lines) or in the region of the shocks as collisions are the main excitation mechanism for strong optical lines. For the lower figure, SNRs are typically located in the shocks region. The sulphur BPT diagram can classify SNRs according to the different ambient densities ("MHD25"+ "HD25" vs "MHD50"+"HD50") as was shown with k-means.}
    \label{fig:bpt_all_data}
\end{figure}

The BPT diagram is a diagram based on strong optical line ratios (typically, $\rm [\ion{O}{iii}]/H\beta$ and $\rm [\ion{N}{ii}]/H\alpha$). It helps to classify the dominant mechanism of ionisation in the observed object (e.g.~an individual object like a SNR or in the whole galaxy).
Two reference lines \citep{Kewley2001, Kauffmann2003} (defined by theoretical modelling and SDSS catalogue analysis \citealt{SDSS2000}) separate objects which are ionised by hot stars (star-forming emission or \ion{H}{ii} regions, lower left part) or by hard radiation of shocks (upper right part).
Mostly, SNRs are located between the theoretical lines (so-called "mixed region") or in the shock-dominated region.
The BPT diagram for our data set ("MC1$_{\rm att}$") is shown in the upper row in Fig.~\ref{fig:bpt_all_data}.
Colours indicate different data sets: "MHD25", "HD25" - blue; "MHD50", "HD50" - red.
The shape of the markers shows the presence or absence of a magnetic field: HD - circles, MHD - stars.
The mean values for different simulations of SNRs start at $t=0.1$\,Myr at the border of the "mixed region", or at the lower limit of the shocks region, then at $t=0.2$\,Myr and move up from the classification line as the gas cools down and shocks start to be observable in the optical band.
Finally, at $t=0.4$\,Myr, shocks are mostly dissipated, optical line ratios have become weaker, and SNRs move back to the mixed region.
This is a typical SNR evolution on the BPT diagram.
We can conclude that this BPT diagram does not reveal different evolutionary paths for our data set.

As was concluded in the PCA analysis above, the [\ion{S}{ii}]($\lambda\,6731$)/$\mathrm{H\alpha}$ line ratio is the most important to detect SNRs and trace the various initial conditions (e.g.~density distribution) at the SNR site.
Therefore, we also plot a sulphur BPT diagram in Fig.~\ref{fig:bpt_all_data}, bottom row.
The colours and symbols are the same as in the classical BPT diagram and the separation line is taken from \citet{Kewley2001}.
During the evolution of the SNR, our mean value is clearly in the shock-dominated region, apart from a few points in the photoionisation region.
The sulphur BPT diagram is sensitive enough to distinguish between the different ambient density distributions at the site of the SNRs.
That is why at any moment of the evolution, our line ratios are divided into two groups: "MHD25"+"HD25" and "MHD50"+"HD50".
The [\ion{S}{ii}]($\lambda\,6731$)/$\mathrm{{H \alpha}}$ line ratio is also very similar for these two groups of simulations and almost does not change.
Due to the known dependence of the [\ion{S}{ii}] doublet on the electron ambient density \citep{Smith1993, Draine2011}, the similarity of the line ratios could be a consequence of the ambient media distribution at the SNR site.
This is discussed further in Section~\ref{sec:results}.

\section{Results}\label{sec:results}
\subsection{Different distances of SNe to MC}
\begin{figure}
    \includegraphics[width=\columnwidth]{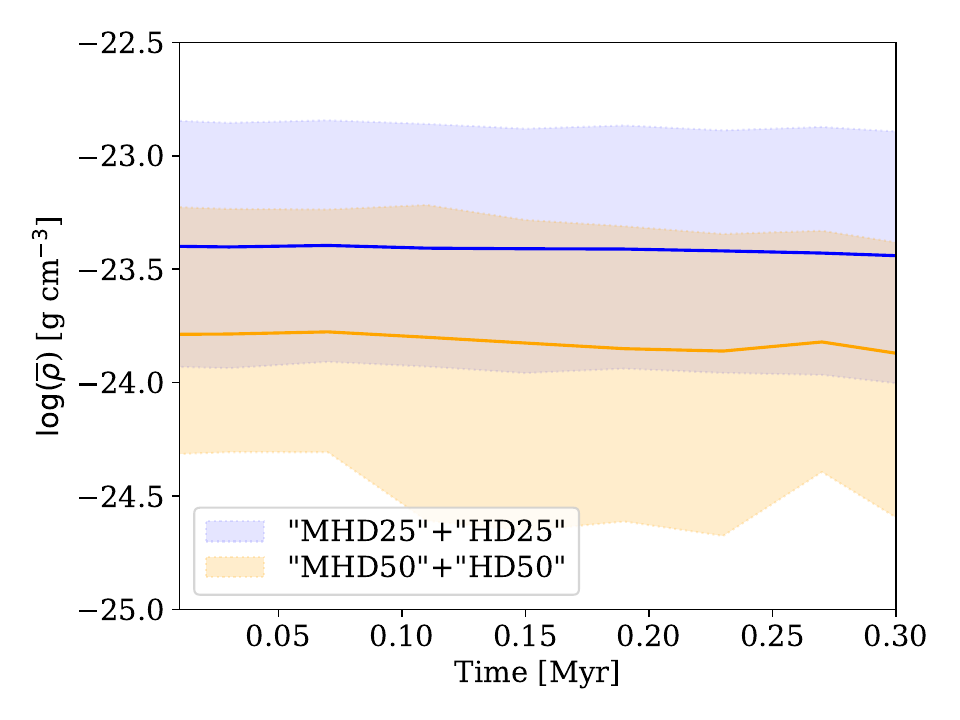}
    \caption{Evolution of the mean density of the ambient medium for the "MHD25"+"HD25" (blue) and "MHD50"+"HD50" (orange) data sets (the mean of means for both data sets). Note that the mean is originally calculated in linear space over each set and presented in log space in the figure. The shaded region shows the interquartile range (25th and 75th percentiles of the data). The "MHD25"+"HD25" data set has overall higher values (around 0.4 orders of magnitude) than the "MHD50"+"HD50" data set.}
    \label{fig:amb_density_25vs50}
\end{figure}

To test the influence of the initial density distribution on the optical emission of a SNR, two positions of the SN relative to the centre of the MC were considered.
The first is located at 25\,pc, and the second one at 50\,pc.
The PCA and t-SNE algorithms both show the statistical difference between these two cases, as we can see from the following k-means clustering and the Rand index in Fig.~\ref{fig:randt_ind}.
The reason for this is the presence or absence of a denser medium near the site of the SN explosion.
For strong optical emission in forbidden lines, a fairly dense environment which will be compressed by the shock wave is required, which will then radiate at low energies of the optical range in the late stages of the evolution of the SNR.
This could cause the rise of optical emission in 25\,pc simulations at a different time as each evolutionary stage of the SN strongly depends on the surrounding medium.

To explain why we see a statistical difference between the 25\,pc and 50\,pc simulations, we performed an analysis of the surrounding (ambient) ISM, as detailed below.
First, we defined the shock front (forward and reverse, but only forward shock detection was used) of the SNR using a shock finding routine based on \citet{Lehmann2016}.
This allows us to determine the shock cells based on the conditions of velocity divergence and density gradient.
Second, we cast six rays starting from the SN explosion position to locate the position of the primary shock with respect to the explosion centre.
It is then possible to mask the SNR bubble, and calculate the median density and interquartile range only of the ambient (unshocked) medium.
Taking into account the time evolution of the media surrounding our SNR (0.4\,Myr), we can check how different the distribution of the medium is at 25\,pc compared to 50\,pc.
The result of this procedure for the whole data set is shown in Fig.~\ref{fig:amb_density_25vs50}.
The mean ambient density for "MHD25"+"HD25" (blue) is, on average, higher than for "MHD50"+"HD50" (orange), as can be inferred from the evolution of the mean and interquartile range around 0.4 orders of magnitude.
Although this difference does not appear significant, a difference is seen for optical radiation.
Optical emission is brighter in older supernova remnants (where the shock wave velocities are usually less than 200\,km\,s$^{-1}$).
This brightness is determined by the density of the ambient medium encountered by the shock front.

If our SNe are placed at different positions in the MC they would encounter different ambient densities that would not necessarily depend on their radial distance from the MC centre.
We would be able to differentiate them only in the case of different mean densities, for example as in the cases of 25\,pc and 50\,pc in our simulations.

\subsection{MHD vs HD runs}
The presence of a magnetic field at the explosion site does not directly affect the optical emission from the supernova remnant.
However, the magnetic field influences the morphological evolution of the SNR (as well as the shock waves).
Moreover, due to compression, the magnetic field strength at the rim of the SNR bubble grows from 4 to 100\,$\mu$G.
Singly ionised particles (S$^{+}$, N$^{+}$) are usually formed in approximately the same temperature zone behind the shock wave.
If there are any thermal instabilities combined with a magnetic field or other shock wave parameters favouring regions where hydrogen is ionised (even when N$^{+}$ and S$^{+}$ are able to recombine) the optical line ratios may change.
Therefore, the difference in the optical line ratio depends on the initial conditions and magnetic field presence.
This can be noted from the grid of shock models calculated using MAPPINGS V \citep{Allen2008}.

We have investigated if the magnetic field (initial condition B-field of 4\,$\mu$G) influences the optical line ratios.
For all simulations, we have two simulation runs -- with (MHD) and without the magnetic field (HD).
Studying the optical lines we could not confirm any statistically significant deviation depending on the magnetic field.
Thus, we show that the optical line ratios do not depend on the magnetic field during the SNR stage to a large extent, or the strength of the magnetic field should be significantly higher.
The magnetic field is also expected to be much more important for high-energy bands (e.g. UV, X-ray, $\gamma$-ray).

\subsection{Attenuation effect}

\begin{figure*}	
    \includegraphics[width=17cm]{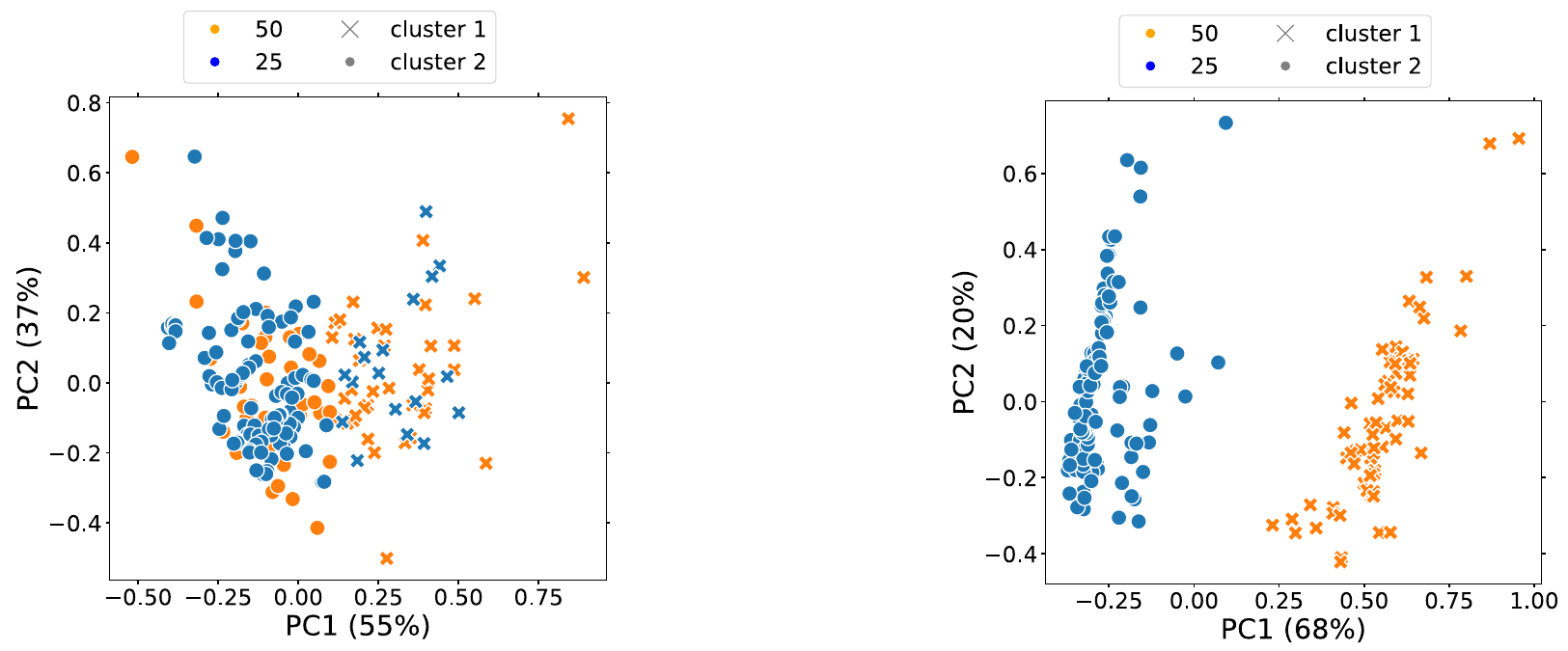}
    \caption{Clustering of all data ("MC1") with k-means after PCA algorithm for line ratios without attenuation ("MC1$_{\rm unatt}$", left panel) and with attenuation ("MC1$_{\rm att}$", right panel). Colours represent real data sets, symbols represent predicted data sets. The best result shows the line ratios with attenuation (right panel). There we can separate the different distances of SN explosion to the centre of mass of the MC.}
    \label{fig:four_panels_plot}
\end{figure*}

\begin{figure}	
	\includegraphics[,width=\columnwidth]
 {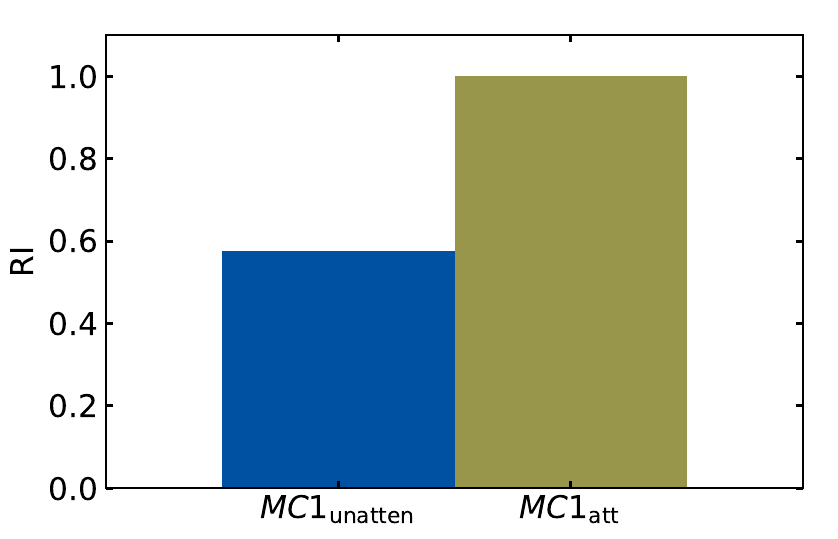}
    \caption{Rand index value ($y$-axis) for unattenuated ("MC1$_{\rm unaten}$") and attenuated ("MC1$_{\rm att}$") line ratios from Fig.~\ref{fig:four_panels_plot}. The line ratio with attenuation (green) shows the best Rand index.}
    \label{fig:rand_index_four_panels}
\end{figure}
\citet{Makarenko2023} demonstrated that attenuation within the SNR cube should be taken into account, as it changes the classification of an object as an SNR based on the optical line ratios.
In this work, we see a similar pattern.
In Figs.~\ref{fig:four_panels_plot} and \ref{fig:rand_index_four_panels}, attenuation plays a crucial role in the search for the effect of a magnetic field or the distribution of matter at the site of an SNR on optical emission.
To check that the [\ion{S}{ii}]($\lambda\,6731$)/$\mathrm{{H \alpha}}$ line ratio is a reliable tracer of the local properties at the SNR site, we need to be sure that attenuation is not mimicking some form of noise for the unattenuated optical lines brightnesses.
We test this by applying random noise to the unattenuated brightness values, at a level corresponding to the range of attenuation between its maximum and minimum values (approximately 80\% and 20\%, respectively; see Paper I, \cite{Makarenko2023}, figure~7), to examine the edge case. Therefore, we do not use the average values provided in Table~\ref{tab:lines_atenuation}.
We repeat this analysis 1000 times.
As a result, we can obtain point separations, for example, on a sulphur BPT diagram in Appendix~\ref{D:bpt+noise}, Fig.~\ref{fig:bpt_s2_noise} using noise.
This means that the [\ion{S}{ii}]($\lambda\,6731$)/$\mathrm{{H \alpha}}$ line ratio is not a universal determinant of the presence of shock waves from an SNR in the environment, and can be confused for other effects.
Therefore, we advise caution when using this type of BPT diagram to make conclusions about correlations of the line ratios with initial conditions at the site of the SNR.

\section{Conclusions}\label{sec:conclusions}
In this paper we analyse optical emission from the SILCC-Zoom simulation data set of SNRs interacting with MCs with and without magnetic fields.
We have a data set of 22 simulations that we post-processed using the CESS package.

To perform a statistical study, we further use the PCA algorithm to pre-process the data, and cluster it using the k-means algorithm.
The Rand index allows us to compare how well the initial labels of the data set fit with the clusters obtained without any prior information on the data.
The Rand index has a value of 1 for the predicted different positions of the SNe with respect to MCs (data set MC1$_{\rm att}$) which means that our initial classification matches the clustered data.
This means that in our simulations we can distinguish different distances to the centre of the MC (25\,pc and 50\,pc) due to the difference in the mean ambient medium density at the site of the SN explosion.
As there is no universal trend of density with radius in MCs, it is not possible to link optical emission with distance from the MC centre. 
Therefore, the mean ambient density at the site of the SN is a relevant quantity in determining subsequent optical emission.
The Rand index shows no statistically significant differences between the simulations with or without a magnetic field, so we can not distinguish them.
Analysing the nitrogen BPT diagram of the optical emission, all our SNRs are mainly located in the mixed region, and the variation between various simulations is minimal.
For the sulphur BPT diagram, we see the same division of the data set as in our clustering analysis, depending on the ambient density distribution at the SN explosion site.
Due to that, we perform an independent analysis of unattenuated optical line luminosity adding random noise.
We show that we can mimic the results of the sulphur BPT diagram with attenuation using unattenuated optical lines and some random noise.
Multi-dimensional analysis of optical emission line ratios does not give extra information about the environmental conditions of the SNR. 
Therefore, we propose to not blindly trust the optical line diagnostic as a probe for the environment near the SNR and as a classification tool for the SNRs.

Nevertheless, we can conclude that realistic modelling of the ISM is an essential component of SNR modelling.
The density distribution at the explosion site will affect not only the attenuation of optical emission, but also the rate of evolution of the SNR.
This leads to different amounts of optical emission throughout the simulations as was shown for a single SNR in Paper I \citep{Makarenko2023}.
Finally, the use of statistical analysis of a large data set is necessary at present.
This allows a less biased assessment of the importance of different parameters (density distributions and magnetic field) on the optical emission of SNRs.

\begin{acknowledgements}

We thank the anonymous referee for constructive comments which helped us to improve the manuscript.
E.I.M. S.W. and D.S. acknowledge the support of the Deutsche Forschungsgemeinschaft (DFG) via the Collaborative Research Center SFB 1601 ‘Habitats of massive stars across cosmic time’ (subprojects B1, B4 and B5).
The following {\sc Python} packages were used: {\sc NumPy} \citep{Numpy}, {\sc SciPy} \citep{SciPy}, {\sc Matplotlib} \citep{matplotlib}, {\sc YT} \citep{yt}, {\sc scikit-learn} \citep{scikit-learn}, {\sc pandas} \citep{pandas}. 

\end{acknowledgements}

%
%

\bibliographystyle{aa}

\begin{appendix}
\section{PCA analysis for MHD vs HD}\label{A:PCA_for_MF}
\begin{figure}
    \centering
    \includegraphics[width=0.8\columnwidth]{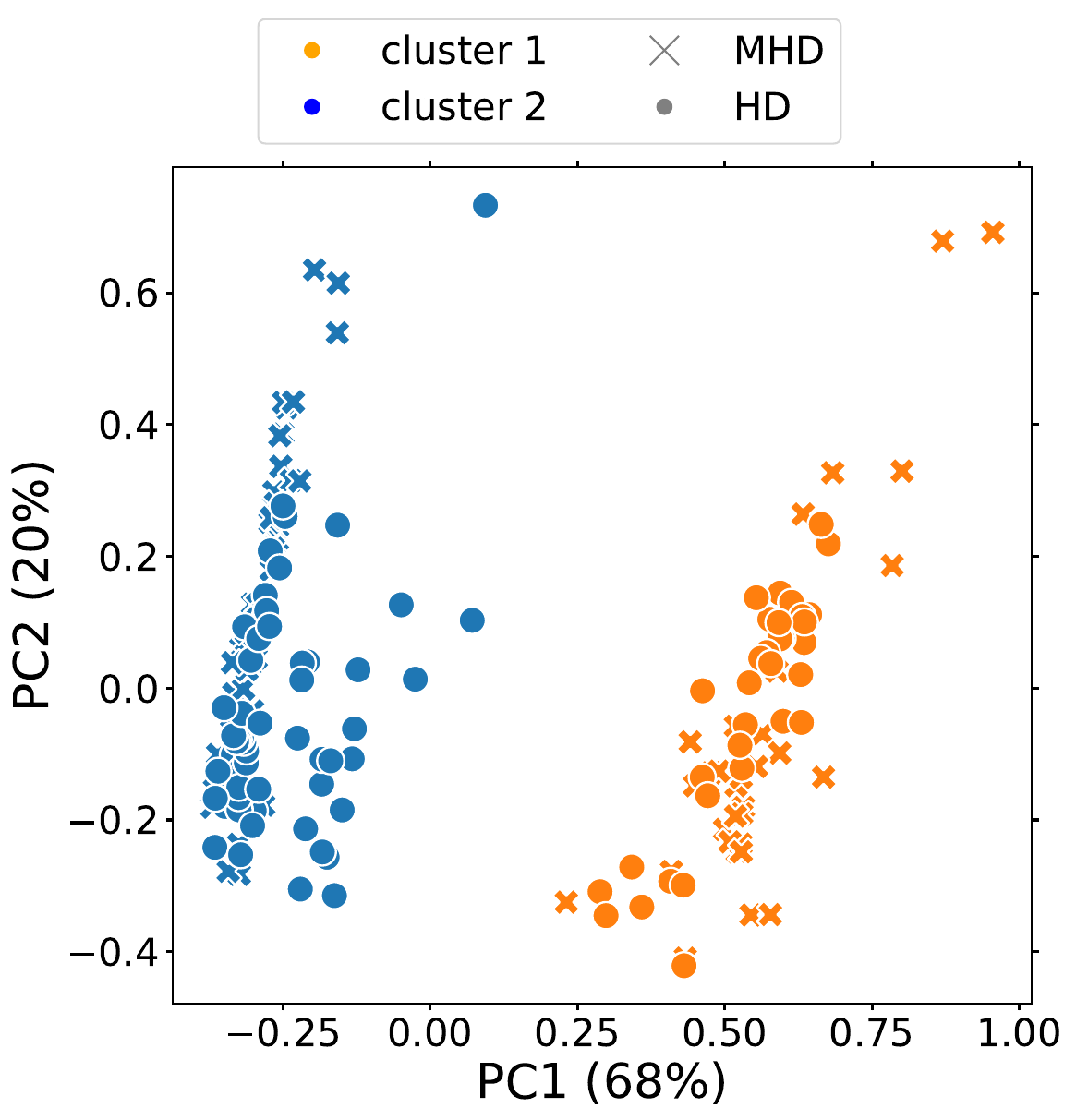}
    \caption{"MHD25", "MHD50", "HD25", "HD50" data sets (e.g. "MC1$_{\rm att}$") after using the PCA algorithm. "MHD" simulations are marked with crosses, and "HD" simulations with circles. The predicted clusters are marked with blue and orange. The higher the percentage for each PC the higher the relative variance in the data set that is observed in the direction of the corresponding eigenvector (for the absolute values see Table~\ref{tab:PC1_PC2_combination}). We use $n_{\rm clust} = 2$ in the k-means algorithm. The data is clearly separated into two distinct clusters by distance: 25\,pc and 50\,pc (see Fig.~\ref{fig:all_data_2or4clust}), while MHD and HD labels are mixed.}
    \label{fig:MHDvsHD_PCA}
\end{figure}

In Fig.~\ref{fig:MHDvsHD_PCA} we present the results of the PCA algorithm for the whole data set ("MC1$_{\rm att}$") divided by k-means in two clusters (n$_{\rm clust}$ = 2).
We add labels to identify simulations with (MHD) and without magnetic field (HD).
Note that the labels are not used in the PCA or k-means, we use them only to check the classification of the data set.
The data is clearly separated into two distinct clusters, but not by magnetic field presence (the colours do not match the shapes, as described in the figure caption). 
Therefore we can conclude that we can not distinguish between simulations with and without magnetic field. 
For the Rand index (see Fig.~\ref{fig:randt_ind}) it reaches a value of 0.5 which also means that the labels do not match the resulting clusters (for non-normalised Rand index).

\section{t-SNE statistical analysis}\label{B:t-SNE}
\begin{figure}
    \includegraphics[width=0.8\columnwidth]{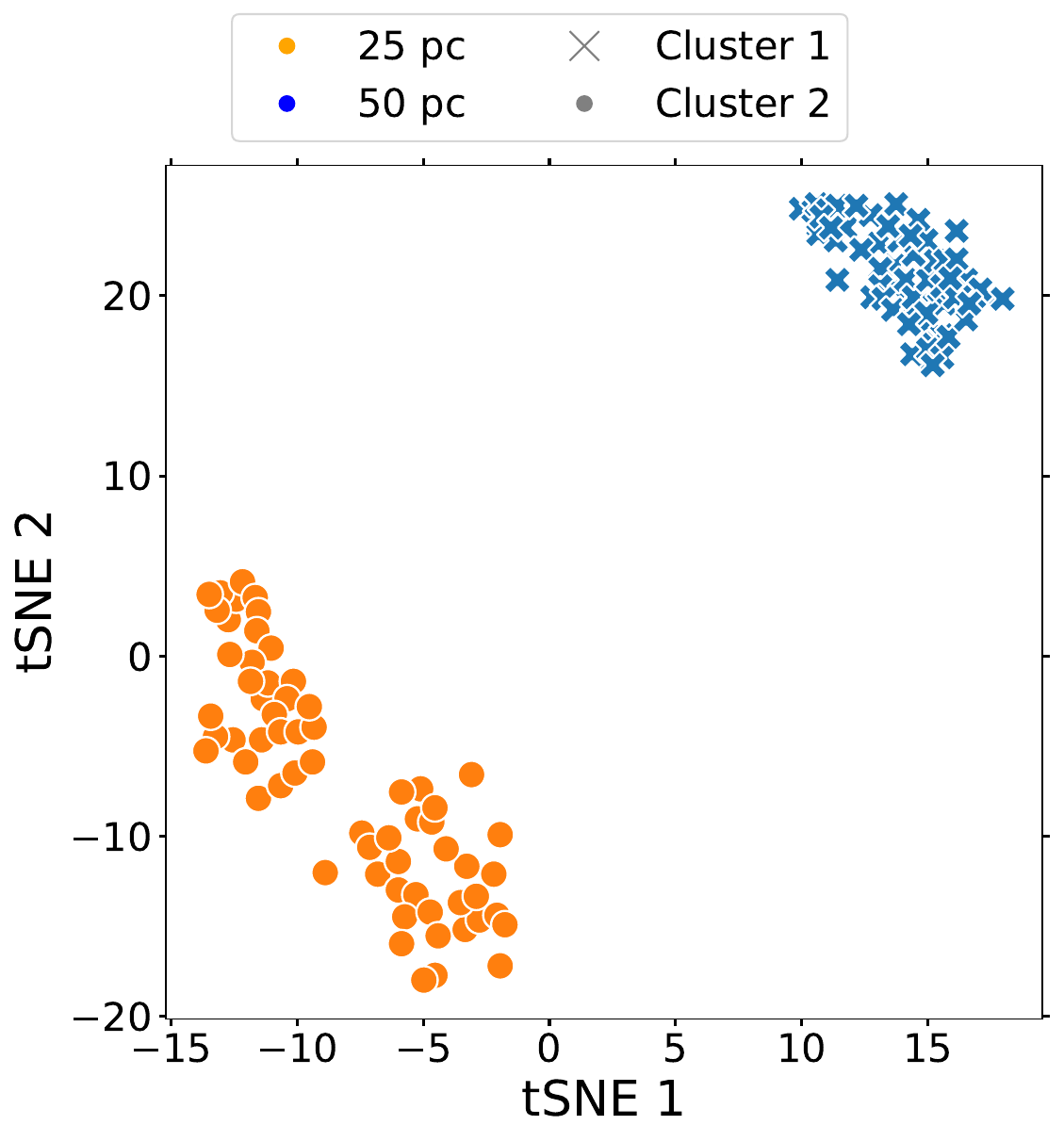}
    \caption{MHD25, MHD50, HD25, HD50 data sets (or "MC1$_{\rm att}$") after using the t-SNE algorithm. We use $n_{\rm clust} = 2$ in the k-means algorithm. Even though three clusters can be visually distinguished, this separation is not related to the presence or absence of a magnetic field. It is also not related to other physical parameters that we might associate with each cluster. That is why, the data is separated into two distinct clusters as for the PCA algorithm for this data set.}
    \label{fig:TSNE}
\end{figure}

Non-linear algorithms such as t-SNE for the dimension reduction of the data set perform better than the linear ones (as PCA), especially for the preservation of local structures (e.g. clusters) of data.
However, for our data set, there was no difference between these algorithms. 
An example of t-SNE and k-means algorithm for the "MC1$_{\rm att}$" data set is shown in Fig.~\ref{fig:TSNE}. 
We can define the same clusters as with the PCA algorithm in Fig.~\ref{fig:all_data_2or4clust}. 
Therefore, we use in this work PCA as it is computationally less expensive.

\section{3-D representation of the statistical analysis (PCA)}\label{C:3DPCA}
\begin{figure}	
	\includegraphics[trim={3cm 0.9cm 1cm 1.6cm},clip,width=1\columnwidth]{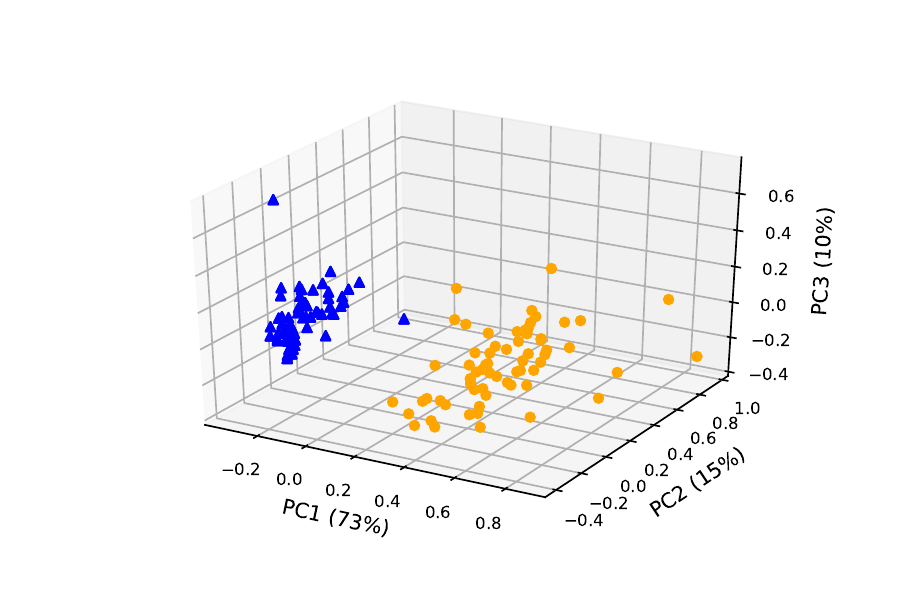}
    \caption{3-D representation of the data "MHD25"+"HD25" (blue) and "MHD50"+"HD50" (orange) after the PCA step. The predicted clusters are marked with circles and triangles. The contribution of each principal component (PC) is as follows: 73\% for PC1, 15\% for PC2, and 10\% for PC3. 
    As in Fig.~\ref{fig:all_data_2or4clust}, we can still clearly see two clusters. However, visually, PC3 does not allow more efficient separation of clusters. Thus, in this work, 2-D visualisation was mainly used (only the two PCs).}
    \label{fig:example_pca_3D_25-50}
\end{figure}

In this work, we used a 2-D representation of the data after the PCA analysis (i.e. we took only the two most important components).
Here we show the 3-D distribution of the data (three principal components) in Fig.~\ref{fig:example_pca_3D_25-50}.
The third dimension contributes little to cluster classifications (only 10-13\%).
In addition, the third dimension is not helpful for better-separating points in space, therefore a 2-D representation is optimal for the k-means clustering.

\section{Sulphur BPT diagram with noise}\label{D:bpt+noise}
\begin{figure}
    \centering
    \includegraphics[width=1\columnwidth]{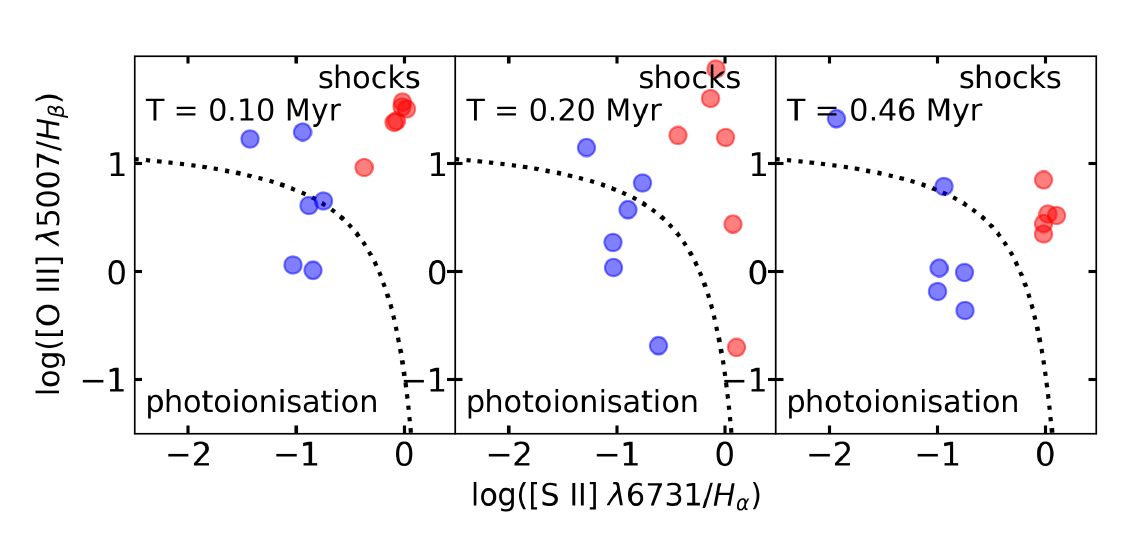}
    \includegraphics[width=1\columnwidth]{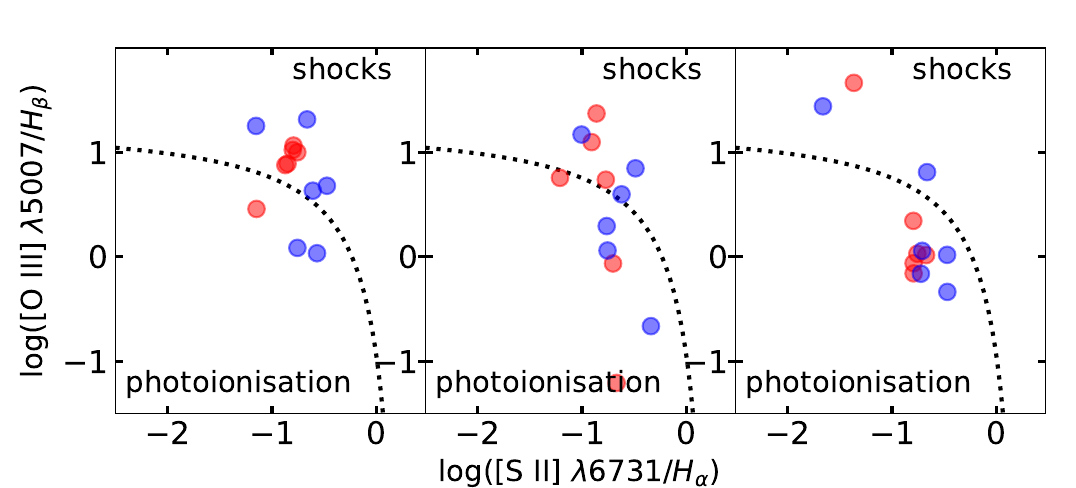}
    \caption{Sulphur BPT diagram for unattenuated line emission with added random noise. Colours are the same as in Fig.~\ref{fig:bpt_all_data}. On the lower panel, noise is higher than the real attenuation percentage (around 80\%~$\pm$~10\%, see Table~\ref{tab:lines_atenuation}). On the upper panel, noise is lower than the calculated attenuation (around 20\%~$\pm$~10\%). The upper panel's circles can be well separated during the whole time evolution (from left to right).}
    \label{fig:bpt_s2_noise}
\end{figure}

To assess whether [\ion{S}{ii}]($\lambda\,6731$)/$\mathrm{{H \alpha}}$ is reliable to identify different ambient density distributions at the SNR site, we plot the sulphur BPT diagram for the unattenuated optical emission lines with added random noise.
Typically, [\ion{S}{ii}]($\lambda\,6731$)/$\mathrm{{H \alpha}}$ is widely used to classify an object as an SNR as it is relatively strong and easy to observe.
However, we found that attenuation plays an important role in this line ratio and can be confused with random noise as in the upper panel of Fig.~\ref{fig:bpt_s2_noise}.

\end{appendix}

\end{document}